# Oxygen Hole Formation Controls Stability in LiNiO$_2$ Cathodes: DFT Studies of Oxygen Loss and Singlet Oxygen Formation in Li-Ion Batteries

A. R. Genreith-Schriever[1,5], H. Banerjee[1,2,5], A. S. Menon[3,5], E. N. Bassey[1,4,5], L. F. Piper[3,5], C. P. Grey[1,5*], A. J. Morris[2,5*]

[1]Yusuf Hamied Department of Chemistry, University of Cambridge, Cambridge, United Kingdom
[2]School of Metallurgy and Materials, University of Birmingham, Birmingham, United Kingdom
[3]WMG, University of Warwick, Coventry, United Kingdom
[4] Materials Department and Materials Research Laboratory, UC Santa Barbara, Santa Barbara, United States
[5]The Faraday Institution, Harwell Science and Innovation Campus, Didcot, United Kingdom

*e-mail: cpg27@cam.ac.uk, a.j.morris.1@bham.ac.uk

## Abstract

Ni-rich cathode materials achieve both high voltages and capacities in Li-ion batteries but are prone to structural instabilities and oxygen loss via the formation of singlet oxygen. Using *ab initio* molecular dynamics simulations, we observe spontaneous O$_2$ loss from the (012) surface of delithiated LiNiO$_2$, singlet oxygen forming in the process. We find that the origin of the instability lies in the pronounced oxidation of O during delithiation, *i.e.*, O plays a central role in Ni-O redox in LiNiO$_2$. For LiNiO$_2$, NiO$_2$, and the prototype rock salt NiO, density-functional theory and dynamical mean-field theory calculations based on maximally localised Wannier functions yield a Ni charge state of *ca.* +2, with O varying between –2 (NiO), –1.5 (LiNiO$_2$) and –1 (NiO$_2$). Predicted XAS Ni *K* and O *K*-edge spectra are in excellent agreement with experimental XAS spectra, confirming the predicted charge states. The calculations also show that a high-voltage O *K*-edge feature at 531 eV previously assigned to lattice O-redox processes could alternatively arise from O-redox induced water intercalation and O–O dimer formation with lattice O at high states of charge. The O$_2$ surface loss route observed here consists of 2 surface O$^{\bullet-}$ radicals combining to form a peroxide ion, which is oxidised to O$_2$, leaving behind 2 O vacancies and 2 O$^{2-}$ ions: effectively 4 O$^{\bullet-}$ radicals disproportionate to O$_2$ and 2 O$^{2-}$ ions. The reaction liberates *ca.* 3 eV per O$_2$ molecule. Singlet oxygen formation is caused by the singlet ground state of the peroxide ion, with spin conservation dictating the preferential release of $^1$O$_2$, the strongly exergonic reaction providing the free energy required for the formation of $^1$O$_2$ in its excited state.

Keywords: LNO, oxygen loss, O redox, singlet oxygen, DFT, DMFT, AIMD simulations, XAS, water.

# 1 Introduction

Ni-rich layered oxides exhibit excellent performance as high-voltage cathode materials, enabling batteries with high energy densities, and are widely used in current electric vehicle batteries.[1,2] They are, however, prone to structural instabilities and degradation involving oxygen loss;[3-6] the degradation is particularly pronounced as the Ni content increases, hampering the development of Co-free next generation cathode materials. Singlet oxygen $^1O_2$ has been captured experimentally[7] or deduced from the large amount of electrolyte oxidation products that accompany electrode degradation, which are unlikely caused by reaction with unreactive triplet oxygen.[8,9] Neither the reason for singlet formation nor its fate after evolution from the surface are, to date, well understood.

At the heart of the O-loss problem is the role that O *vs.* Ni redox plays in the partially reversible removal and reinsertion of Li. The redox activity of LiNiO$_2$ (LNO) is generally attributed predominantly to Ni, *i.e.*, it is assumed that in the process of charging and discharging the battery, it is mainly Ni that is oxidised and reduced.[5,10-13] Within the formal oxidation state model, Ni is considered to be oxidised from +3 to +4 upon delithiating LNO.[5,13] It is clear that this ionic picture does not fully capture the nature of the Ni–O bonds which are known to exhibit covalent character with hybridisation between the Ni *d* and O *p* states. The charge states of both species will therefore deviate from their formal oxidation states.[10,14] However, the consensus is generally that Ni is not solely oxidised/reduced and that while O is somewhat involved in the Ni–O redox processes, Ni activity remains dominant. In agreement with this,[5,10,11] X-ray spectroscopy (XAS) of Ni-rich layered oxides shows a shift of the Ni *K* edge on delithiation towards higher energies,[15-17] which is commonly interpreted as experimental confirmation of the changes in the Ni charge state. Kong *et al.* have previously reported partial O oxidation upon delithiation of LNO but continue to attribute the main redox activity to Ni.[10] In a similar vein, Korotin *et al.*[18] and Foyevtsova *et al.*[19] invoke ligand holes on oxygen, suggesting a partial charge transfer of oxygen to Ni, but simultaneously propose a disproportionation of the Ni charge states, which has not yet been confirmed experimentally. A greater involvement of O is considered in O redox materials, where O redox goes beyond $\sigma$(Ni–O) interactions, *e.g.*, in Li-excess cathode materials,[14,20-23] in the form of orphaned O states[14] or stabilising delocalised metal-oxygen $\pi$ interactions.[20] Little work to date has considered the role that the rehybridisation of Ni-O bonds plays in oxygen loss in non-Li-excess layered oxides.[17,24,25] This is the focus of this work.

The charge states in the LNO system have not previously been investigated in depth, especially as there are numerous technical challenges regarding the quantification of species in hybridised states. Wave-function based charge analyses proposed, for example, by Mulliken[26] partition the wave function according to atomic orbitals, which is a questionable approach for characterising hybridised states. Determining oxidation states using density-functional theory (DFT) calculations requires the ground-state electron density to be partitioned and assigned to the individual ions; the resulting ionic charge therefore often depends on the choice of partitioning scheme, the most established being Bader charge analysis.[27] Reeves and Kanai[28] as well as Quan and Pickett[29] suggest that projecting the DFT charge density onto maximally localised Wannier functions yields the most reliable charges of atoms in molecules[28] and ions in crystals.[29] This is the approach that is applied in this work.[29]

To understand the role of the surface itself in singlet oxygen formation, Wandt *et al.* have studied thermally induced $O_2$ evolution from Ni-rich layered oxides into the gas phase. They report a rise in $^1O_2$ evolution around 500 K,[7] confirming the existence of at least one route for $^1O_2$ formation without any electrolyte. Houchins *et al.* have recently proposed that in metal-air batteries, singlet oxygen forms via a disproportionation reaction of superoxide radicals in solution.[30] As the superoxide radicals are part of the metal-air electrochemistry and they play no role in Li-ion batteries, it is unlikely or at least unclear if the formation of singlet oxygen in Li-ion batteries with layered oxide cathode materials follows the same mechanism.

Understanding how oxygen is evolved requires an atomistic examination of the surfaces involved. Based on scanning electron microscopy and X-ray diffraction, Zhu and Chen[31] and Garcia and co-workers[32] showed the prevalent surfaces in Ni-rich layered oxide cathodes to exhibit facets of the (012), (001), and (104) families, in line with computational predictions.[33-35] The facets found to cover the greatest surface area, over a range of particle morphologies, were (012) facets,[31] which are predicted to be O-terminated at synthesis conditions.[33] The O-terminated (012) facet is therefore an ideal model system to explore (singlet) oxygen loss in Ni-rich layered oxides.

Here, we re-examine the classic picture of transition-metal centred redox in stoichiometric $LiNiO_2$ and delithiated $NiO_2$. At high states of charge, LNO is known to decompose forming surface-reconstruction phases with rock salt-like structures,[5,36] the prototype material of which, NiO, is studied here. A variety of computational tools ranging from density functional theory (DFT) to dynamical mean-field theory (DMFT) calculations are employed to determine oxidation states; we use charge-analysis schemes ranging from a Bader charge analysis to the Wannier transformation and integration over the impurity Green's functions within DMFT, an approach that promises to become a powerful tool for analysing charge states in battery materials. We validate our predictions against experiment by predicting the XAS spectra at the Ni *K* and O *K* edges of LNO and $NiO_2$ and comparing them with previously published experimental Ni *K* and O *K*-edge X-ray spectra of $LiNiO_2$ at varying states of delithiation.[17] Employing *ab initio* molecular dynamics (AIMD) simulations, we analyse the spontaneous $O_2$ release from the (012) surface of delithiated LNO and for the first time computationally capture singlet oxygen as it forms at the cathode surface. We propose a comprehensive mechanism for the observed route of oxygen loss and singlet oxygen formation, pivotal for mitigating electrolyte oxidation pathways associated with singlet oxygen release.[37]

## 2 Oxidation states

Even at the most basic level of charge analysis, a Bader partition of DFT densities from PBE+$U$ calculations ($U_{\text{eff}}$ = 6 eV, as proposed by Das et al.[5] and validated with hybrid functionals and dynamical mean-field theory in this study) shows that Ni exhibits a similar charge state in materials that show three different formal Ni oxidation states, NiO, LiNiO$_2$ and NiO$_2$ (see Fig. 1 a, first blue bar in each set). By contrast, Fig. 1 a shows that the O charge state, on the other hand, changes significantly upon delithiation of LiNiO$_2$, and even further upon formation of NiO. The electronic density of states (DOS) (see Fig. 1 b) suggests ferromagnetic behaviour at 0 K. Of particular interest are the states just below the Fermi energy, as these are the states that electrons are removed from upon delithiation of the material. The character of these states can be estimated by projecting the DOS onto spherical harmonics, *i.e.*, local atomic orbitals. They indicate that the states just below the Fermi energy have predominantly O (red) contributions and only to a small extent Ni (blue) contributions. The charge density of these states (yellow isosurface in Fig.1 c) confirms pronounced O $p$ contributions and weaker Ni $e_g$ contributions just below the Fermi energy (the states are highlighted in the DOS in Fig. 1 b).

Due to the hybridised nature of the states near the Fermi energy and the challenges in separating the charges, the charges were also analysed at a more advanced level with a wannierisation of the PBE Kohn-Sham orbitals. In the case of LNO, a Ni $d$ wannierisation shows large O $p$ contributions for the nominal Ni $e_{g*}$ states or highest occupied states (orbitals shown in yellow and green in Fig. 1 c), confirming that the Ni and O bands are strongly hybridised. The electronic band structure (see SI) suggests that the Ni $d$ and O $p$ bands are very close in energy, requiring both states to be included in the correlated subspace of the wannierisation. If the O states are explicitly included in the wannierisation using a *d-p* model, Ni $d$ states are localised on Ni centres and O $p$ states on O centres (see Fig. 1 c). NiO$_2$ also requires a *d–p* model, while NiO contains $d$ and $p$ bands that are well separated, making a $d$ model sufficient (see band structure and additional NiO *d–p* analysis in the SI). The resulting ionic charges calculated from Wannier occupancies (see Fig. 1 a, middle bar in each set) have significantly larger values than the Bader charges (by 0.5-0.8 $e$ in the case of Ni and 0.3-0.9 $e$ in case of O), but confirm all trends seen with the Bader analysis of the PBE+$U$ states; again, the Ni charges are nearly identical in LiNiO$_2$, NiO$_2$, and NiO, while the O oxidation states vary substantially, suggesting pronounced O redox activity.

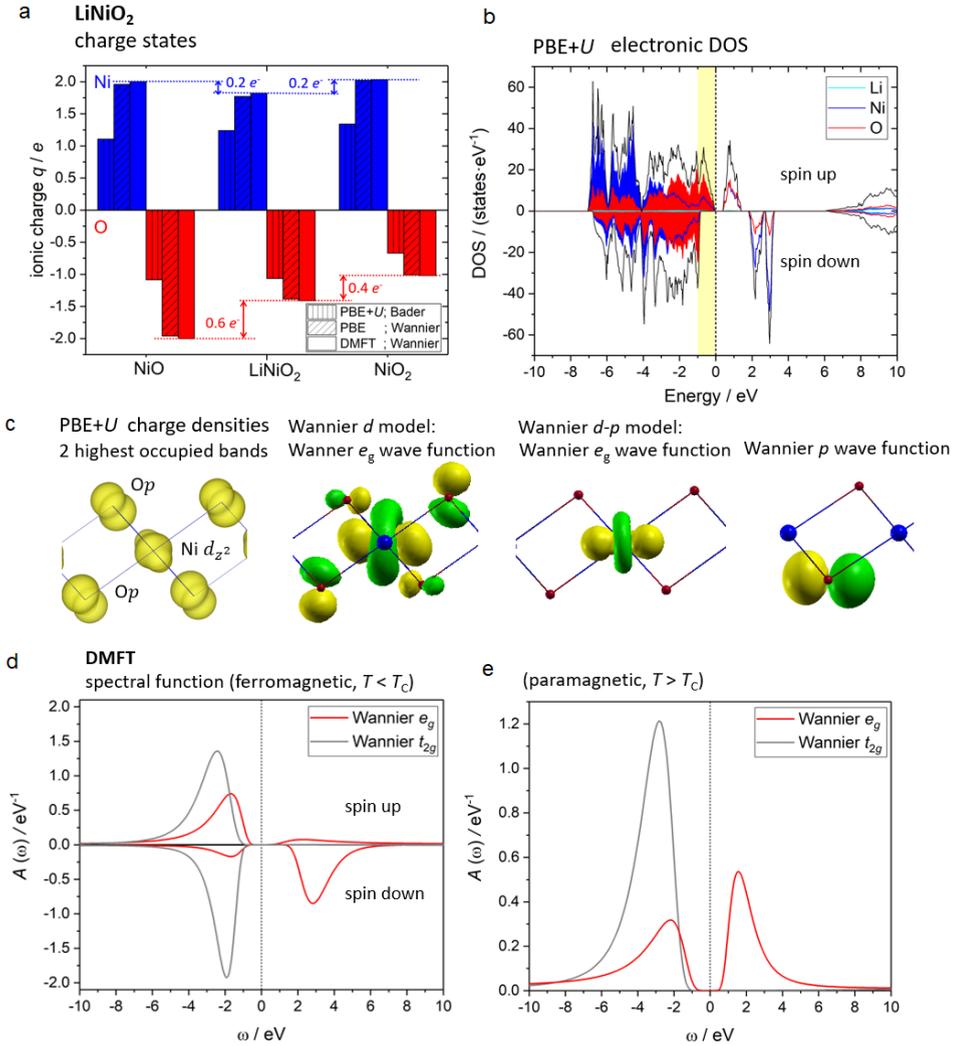

**Fig. 1 Oxidation states in LiNiO$_2$ and related materials and LiNiO$_2$ electronic structure. a**, Ionic charges of Ni (blue) and O (red) in NiO, LiNiO$_2$, and NiO$_2$ determined with PBE+$U$ ($U_{eff}$ = 6 eV) and a Bader analysis (left bar in each set); Wannier occupancies based on PBE Bloch-states (middle bar); full charge self-consistent DFT+DMFT based on wannierisation (right bar). **b**, PBE+$U$ electronic DOS with Ni contributions shown in blue and O contributions shown in red. The highest occupied states of LiNiO$_2$ are highlighted in yellow; they indicate strong O contributions. **c**, the yellow isosurface shows the charge density of the highest occupied states, as obtained from PBE+$U$ calculations for $E_F$ – 1 eV to $E_F$; the yellow and green isosurfaces depict the Wannier wave functions after rotation relative to a global coordinate system. The charge density shows Ni $d_{z^2}$ character and strong O $p$ character. The Wannier $d$ wavefunctions obtained from a Ni $d$ model of LNO show Ni $d_{z^2}$ character and significant contributions from O $p$ states, which are moved to an O centre if the O $p$ states are explicitly included in the active basis set. **d**, the spectral function obtained from a DMFT impurity model of Ni for the ferromagnetic material at $T < T_C$ (O is treated as part of the electron bath in the DMFT calculations).

**e**, spectral function of the paramagnetic material at $T > T_c$ showing good agreement with the small bandgap paramagnetic character seen in experiment. [38-43]

A question then arises as to whether this O character just below the Fermi energy and the resulting redox activity of O are an artefact, *e.g.*, of the DFT calculations themselves or of the static Hubbard $U$ treatment of electron correlation, which lowers the energy of the Ni states, potentially causing unphysical O $p$ character just below the Fermi energy. Dynamical mean-field theory (DMFT) calculations were therefore performed to address this, which account for electron correlations via a mean-field approach and obtain charge states by integrating over the impurity Green's functions in a Wannier basis. The DMFT spectral function is shown in Fig. 1 d for the ferromagnetic case, resembling the DFT DOS at 0 K. At temperatures above the Curie temperature, the material becomes paramagnetic, as shown in Fig. 1 e, where the two $e_g$ states are each half occupied (red line). This small band-gap paramagnetic behaviour is in good agreement with experimental findings at battery operating temperatures.[39-43]

Returning to the question of the charge states of Ni and O, Fig. 1 a shows that the DMFT charges (third bar of each set) are nearly identical to the Wannier charges based on the PBE densities (middle bar of each set). DMFT yields a Ni oxidation state of ∼ +2 in all three materials, LiNiO$_2$, NiO$_2$, and NiO, while the charge of O is *ca.* −1.5 in LiNiO$_2$, −1 in NiO$_2$, and assumes its most stable state of −2 only in the decomposition product NiO.

A standard experimental procedure to track oxidation state changes in battery materials is X-ray absorption spectroscopy (XAS). To validate the computed charge states and directly compare our predictions with experiment, we calculated Ni $K$ and O $K$ spectra for LNO and NiO$_2$ (see Fig. 2 a+b) and compared them with experimental Ni $K$ and O $K$ X-ray spectra of a 2% W-doped LNO cathode at varying states of delithiation (see Fig. 2 c+d).[17] The spectra were collected in the bulk-sensitive fluorescent yield mode.

Fig. 2a shows a clear shift of the edge position in the calculated Ni $K$ spectra – with the calculations yielding a Ni charge state of +2 both in LNO and NiO$_2$. The magnitude of the shift is in excellent agreement with the edge shift observed in experiment (see Fig. 2c). While our DFT calculations find the Ni charge to remain +2, they predict that the local Ni-O bonding environment changes on delithiation (see Fig. 2e). In line with previous computations[5,44] and EXAFS results[45] we find LNO to be Jahn-Teller distorted at zero Kelvin. Ni is coordinated by two O at a longer distance of 2.1 Å and four O at a shorter distance of 1.9 Å. NiO$_2$, by contrast, is undistorted, with Ni being surrounded by 6 O at a distance of 1.9 Å. Leaving aside changes in the symmetry of the bonding environment, the average bond lengths also change on delithiation, decreasing from 2.0 Å in LNO to 1.9 Å in NiO$_2$.

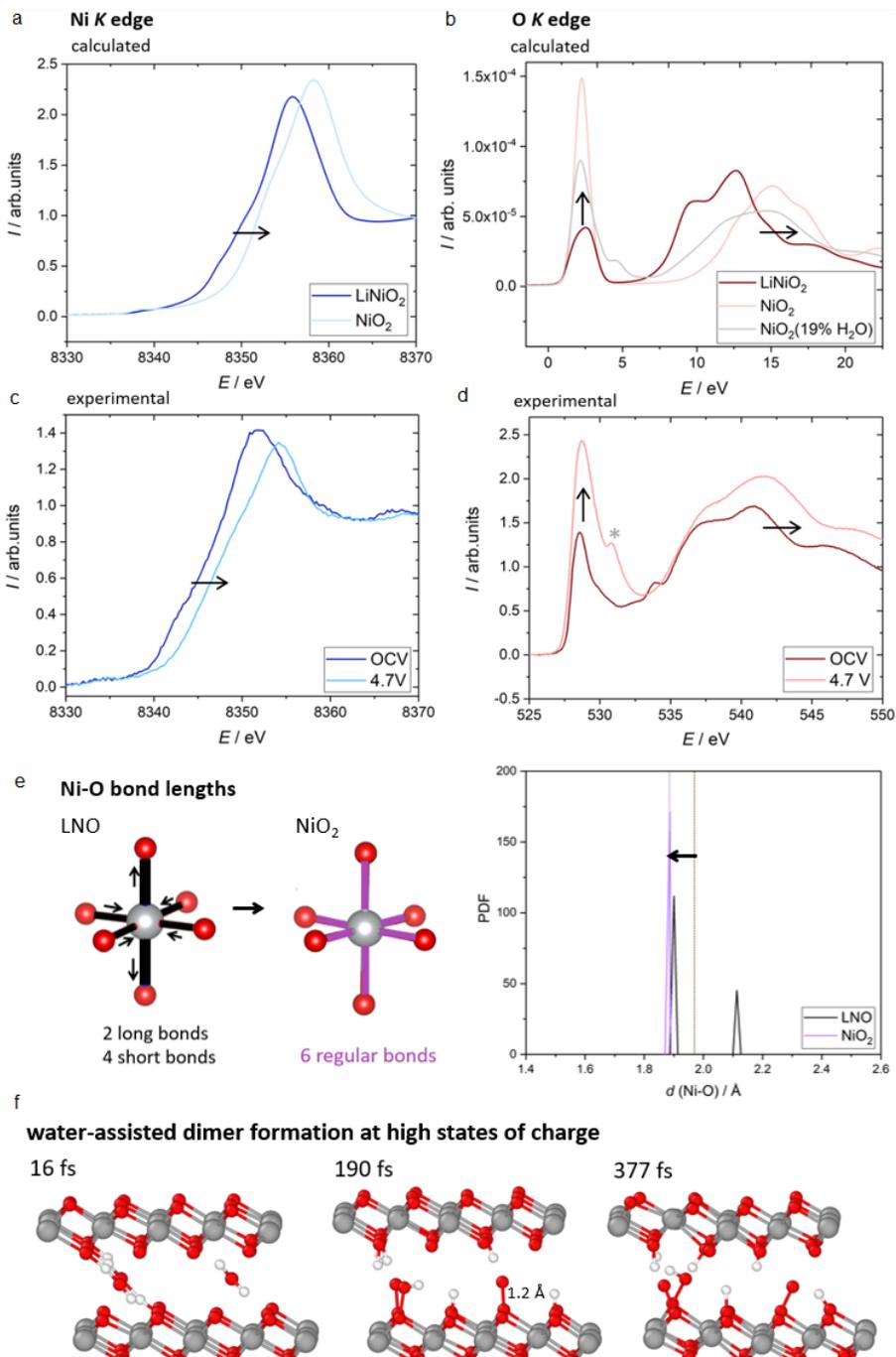

**Fig. 2 X-ray absorption spectra. a**, calculated Ni *K*-edge spectra of LiNiO$_2$ and NiO$_2$. **b**, calculated O *K*-edge spectra of LiNiO$_2$, NiO$_2$, and hydrated NiO$_2$. **c**, experimental Ni *K*-edge spectra of LiNi$_{0.98}$W$_{0.02}$O$_2$ at OCV and charged to 4.7 V. **d**, experimental O *K*-edge spectra of LiNi$_{0.98}$W$_{0.02}$O$_2$ at OCV and 4.7 V. The asterisk

denotes a signal arising at 531 eV at high states of charge, conventionally assigned to lattice O redox. The peak at 534 eV in the OCV data stems from surface carbonate species. **c+d** adapted from Menon et al.[17] **e**, loss of Jahn-Teller distortions and decrease of the mean Ni-O bond length of LNO on delithiation. **f**, water-assisted formation of O-O dimers, as obtained from AIMD simulations at different timesteps. The snapshot at 377 fs was used for the prediction of the XAS spectrum of hydrated $NiO_2$ in **b**.

The O K edge spectrum shows overall greater changes on delithiation than the Ni K spectrum, both computationally (see Fig. 2b) and experimentally (see Fig. 2d). Theory and experiment are again in good agreement regarding the evolution of features, showing a pronounced increase of the O pre-edge peak– a measure of the covalency of the Ni–O bond and the hole concentration of these states– as well as an overall shift of the spectrum to higher energies. An exception is a small additional feature emerging around 531 eV in the experimental O K spectrum at high state of charge that is not seen in the calculated spectrum of defect-free $NiO_2$. This could stem, for example, from additional species such as point defects or additional electronic phenomena beyond the DFT level of the O K spectra calculations. Lee and co-workers have recently proposed an electrochemical co-intercalation of water into $LiNiO_2$,[46] a phenomenon that is well established for layered Na cathodes. The question arises if water could be causing the additional XAS feature. This requires water to generally be available in the system. Given that the reaction of singlet oxygen with battery electrolytes has been reported to produce water[37] and, additionally, protic electrolytes have been found to directly transfer protons to Ni-rich oxide surfaces at high states of charge,[47] water is expected to occur even in nominally water-free setups, suggesting water intercalation is at least plausible. AIMD simulations of hydrated LNO were performed at varying states of delithiation. At high states of charge, water molecules intercalated into $NiO_2$ show the formation of O–O dimers with lattice O ions with a bond length of *ca.* 1.2 Å (see Fig. 2f). Some of the dimers and the nearby lattice O are protonated, with dynamic proton exchange occurring. *N.B.*, the dimers do not leave the lattice sites in the course of our simulations. A systematic investigation of the process and a potential involvement of electronic quasi-particles (see SI) is underway but it is worth noting at this stage, first, that such dimers yield XAS signals matching the additional high-voltage feature seen experimentally (see Fig. 2b+d) and, second, that an XAS spectrum calculated for the hydrated material shows excellent overall agreement with the experimental spectrum at 4.7 V.

## 3 Surface oxygen loss

As LNO is known to decompose through surface reactions, it is pivotal to understand how the bulk oxidation states discussed above compare to oxidation states at the surface. We explore the prevalent (012) facet,[31,32] where Li is extracted and inserted during cycling, and we investigate the fully delithiated, unstable material $NiO_2$ (see Fig. 3 a). Since the analysis of the oxidation states showed that the Bader charge analysis gave the correct relative changes of the oxidation states in bulk $LiNiO_2$ and $NiO_2$ without the necessity to assess the suitability of different Wannier models, the surface charges of the (012) facet of $NiO_2$ were evaluated with a Bader charge analysis of the PBE+$U$ charge density. This shows that O at the surface is oxidised even further than in the bulk (see Fig. 3 a). Each Ni-O layer exhibits O species with two slightly different charge states at the surface, both with a lower electron density than in the bulk.

AIMD simulations of the (012) $NiO_2$ surface show a spontaneous loss of $O_2$ molecules and even the formation of singlet oxygen in the process (see Fig. 3 b) at simulated temperatures from 300 – 800 K. Fig. 3 b shows the free energy profile of an exemplary PBE+$U$ trajectory at *ca.* 450 K, suggesting the reaction releasing one $O_2$ molecule liberates 3.0 eV. This was confirmed with AIMD simulations including dispersion interactions and based on hybrid functionals, which also show spontaneous $O_2$ loss to release *ca.* 3 eV. As seen in Fig. 3 b, the reaction appears barrierless, suggesting that the activation barrier is smaller than the energy fluctuations due to thermal vibrations (<0.05 eV).

At the beginning of the simulation, the O species at the surface of $NiO_2$ are separated by *ca.* 2.8 Å. An analysis of the magnetic moments yields *ca.* 1.1 $\mu_B$ per O (see Fig. 3 b), indicating the presence of unpaired electrons. The magnetisation density shows no unpaired electrons in the bulk, but positive spin density on the surface O that interacts antiferromagnetically with negative spin density on the nearest Ni ions (see SI), suggesting radical character of the surface O, *i.e.*, $O^{\bullet-}$. When lattice vibrations bring 2 $O^{\bullet-}$ radicals close together (*ca.* 1.4 Å) at 369 fs, the electronic character of the surface O changes, causing a steep decrease of the magnetic moments to *ca.* 0 $\mu_B$, indicative of the formation of a peroxide ion. The charge density and electron localisation function show a covalent O–O bond of the dimer (see SI), in agreement with an $O_2$ dimer predicted by Kong and co-workers for a different surface, the (104) facet,[10] which – although it is predicted to be particularly stable – it is found to cover only a small percentage of the surface area of NMC particles.[31] The $O_2$ dimer formation Kong *et al.* report is related to the one observed here, but differs in that it requires oxide ions to combine across $NiO_2$ layers, while in the case of the (012) facet we see the oxide ions combining with oxide ions from the same $NiO_2$ layers, a process occurring at lower temperatures.

Fig. 3 b shows the change in electron numbers throughout the reaction, according to a Bader analysis. The O forming the peroxide are oxidised. While some of the electron density drawn from the peroxide in the oxidation is transferred to the nearest Ni ions, the majority goes to the nearest surface $O^{\bullet-}$ radicals. As the peroxide is oxidised, the O–O distance decreases further to *ca.* 1.2 Å at 384 fs, typical of an $O_2$ molecule. Fig. 3 d shows a closeup of the free energy profile. The two $O^{\bullet-}$ radicals (A) form the peroxide at 369 fs (B), which is oxidised continuously to $O_2$ – in double coordination at 384 fs (C), single coordination at 387 fs (D), and desorbed at 394 fs (E), assuming a Ni–O bond cut-off length of 2.05 Å (based on the charge density and denote positive spin density and cyan isosurfaces negative spin density, isosurface levels of $\pm 5.3 \times 10^{-5}$ $e/Å^{-3}$). **f**, closeup of the magnetic moments of the two O$^{\bullet-}$ forming the O$_2$ molecule as a function of time. The moments exhibit a steep decrease around 369 fs to zero, indicating a singlet ground state of the peroxide ion, and steadily increase from *ca.* 0.3 $\mu_B$ at 371 fs to *ca.* 0.8 $\mu_B$ from 500 fs onwards. Once released from the surface, $^1$O$_2$ is no longer a ground state and looks identical to $^1$O$_2$ obtained for the gas phase ($^1\Delta$, see SI).

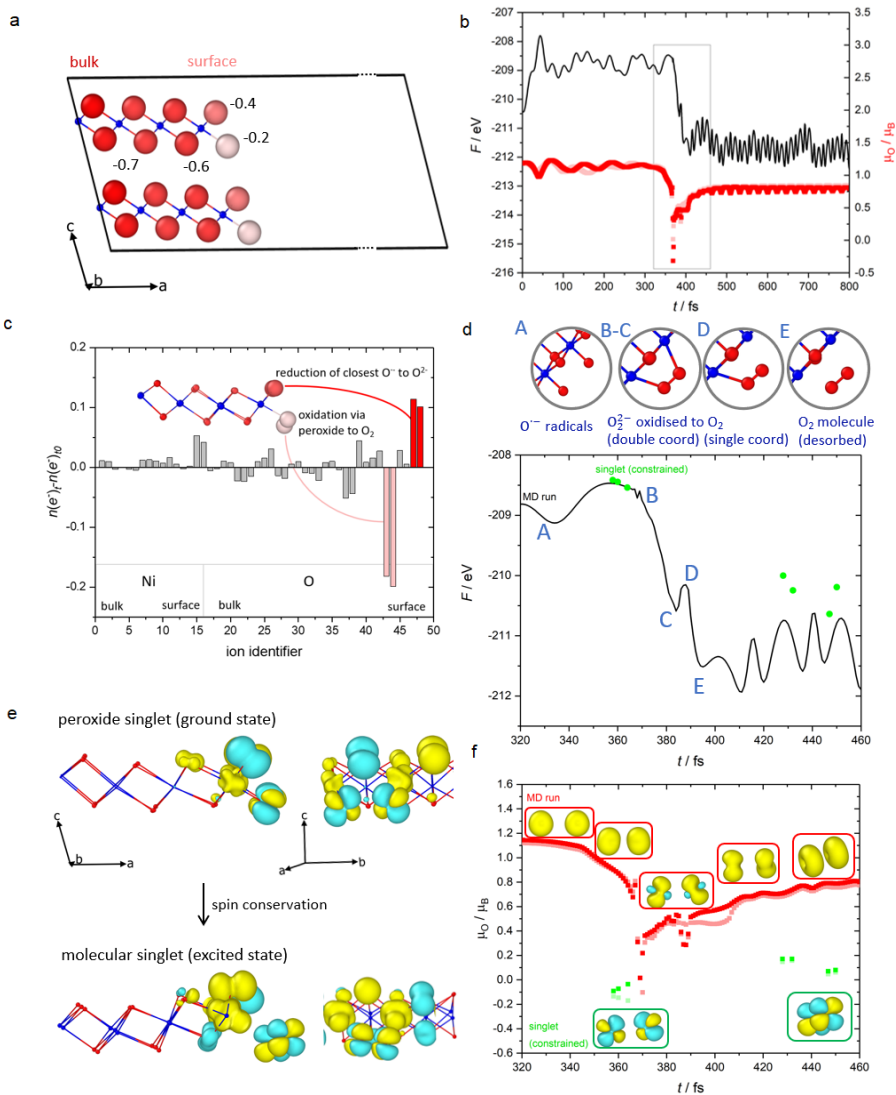

**Fig. 3 Mechanistic analysis of the observed route of O₂ evolution and the formation of ¹O₂. a**, Oxidation states of the O ions near the (012) facet of $NiO_2$ according to a Bader charge analysis of the PBE+$U$ charge density; lighter colours represent higher states of oxidation. Surface O are in a yet more oxidised state than bulk O. **b**, free energy profile (black) and magnetic moments of surface O species (red) of a route of spontaneous O₂ loss, as found in AIMD simulations at ca. 450 K. The reaction occurs around 370-400 fs, where the free energy decreases by ca. 3 eV. The magnetic moments suggest radical character at the beginning of the simulation, $\mu_O \approx 1.1 \mu_B$, and an abrupt change in electronic character during the reaction. **c**, differences in electron numbers of all ions in the simulation cell between 389 fs (when O₂ just desorbed) and the start of the simulation based on a Bader analysis. **d**, closeup of the free energy profile with characteristic points indicated; O•⁻ radicals (A) combine to form a double coordinated peroxide (B), which is continuously oxidised to an O₂ molecule (C-E) and through a single coordinated transition state (D) desorbs as O₂ (E). The charge is transferred to the nearest O•⁻ radicals (see c). **e**, magnetisation density of the peroxide ground-state singlet state and the excited molecular singlet state (yellow isosurfaces

electron localisation function). Most of the oxidation occurs between B and C. The released oxygen molecule vibrates around an average bond length of 1.2 Å and leaves behind two oxygen vacancies and two reduced surface O. The free energy curve shows that *ca.* 1.7 eV (55%) of the free energy gain is due to the peroxide formation and oxidation, while *ca.* 1.3 eV (45%) are due to the remaining oxidation and O$_2$ desorption. The activation barrier of the desorption is *ca.* 0.04 eV – easily overcome by thermal energy, especially in the light of the large amount of energy released through the peroxide formation. The rate-determining step is therefore most likely the peroxide formation (with a negligible barrier at 450 K).

## 4 Singlet oxygen

With this mechanistic understanding of the observed route of oxygen loss from delithiated LNO, we can turn to the question of why singlet oxygen is formed. O$_2$ molecules in the gas phase have a triplet ground state $X^3\Sigma^-_g$ separated by an energy gap of *ca.* 0.9 eV from the first excited state (CASSCF,[48] exp[49]), *i.e.*, the singlet $a^1\Delta_g$ state. Higher in energy still is the $b^1\Sigma^+_g$ state, which is generally not observed to react chemically as it physically deactivates quickly into the $^1\Delta$ state. To unravel the mechanism of the singlet formation in our simulations, it is essential to identify $^1$O$_2$ signatures in the simulations.

The two highest-energy electrons of $^1$O$_2$ are of opposite spin, giving the singlet an overall magnetic moment of zero. AIMD simulations show that at 369 fs the magnetic moments of the two oxide radicals forming the peroxide drop and approach zero briefly (see Fig. 3 b+d) before steadily increasing from *ca.* 0.3 $\mu_B$ (at 371 fs) to *ca.* 0.8 $\mu_B$ (from 500 fs onwards). Albeit briefly, this points towards singlet character of the electronic ground state, and this is the ground state of the peroxide ion. Through constraining the number of unpaired electrons in the simulation cell, we can also enforce the singlet state before the transition, and we see it become degenerate with the AIMD ground state (nearly degenerate *ca.* 15 fs before the drop in the magnetic moments, and fully degenerate *ca.* 3 fs before the drop, see the green data points in Fig. 3 d+f). This singlet state forms as the 2 oxide radicals forming the peroxide come close enough together to start interacting with each other but before any oxidation takes place. The magnetisation density of the singlet state is illustrated in Fig. 3 e+f. Two states with strong O *p* character are seen, one populated with a spin-up electron and the other with a spin-down electron. The lobes of the negative spin state are slightly larger than the lobes of the positive spin state, resulting in a magnetic moment of *ca.* $-0.1\ \mu_B$.

As the simulation proceeds and the peroxide is oxidised, the negative lobes shrink until they disappear, and the magnetic moments increase to $0.8\ \mu_B$ (see Fig. 3 b+f), suggesting triplet character of the ground state $^3O_2$ molecule after it has left the surface. Through constraining the number of unpaired electrons, the singlet state is also obtained for $^1O_2$ after it has left the surface. The magnetisation density seen in Fig. 3 f looks very similar to the singlet seen during the peroxide formation, but in the $^1O_2$ molecule, the planes of the lobes are fully aligned, suggesting a stronger interaction (at a smaller distance; earlier 1.4 Å between the interacting radicals *vs.* 1.2 Å in the molecule). Fig. 3 d shows that the gas phase singlet is no longer degenerate with the ground state but an excited state.

In a DFT gas phase calculation, $^1O_2$ can also be obtained by constraining the magnetic moments of each O atom to zero. The gas phase $^1O_2$ shows the same magnetisation density as the gas phase $^1O_2$ at the surface of $NiO_2$ (see SI). The highest energy level is degenerate across the two spin channels and occupied by one spin-up and spin-down electron each, indicating that this type of singlet seen both in the gas phase and evolving from the $NiO_2$ surface in our simulations corresponds to the $^1\Delta$ singlet where the spins are paired, giving the singlet a closed-shell character.[50,51]

## 5  Discussion

Our results suggest that the redox processes in $LiNiO_2$ need to be reconsidered. While the established ionic model assigns all redox activity to Ni and the established covalent model acknowledges that O is somewhat involved in the redox process, the extent of the O involvement in Ni–O redox seen in Fig. 1 is unexpected and much greater than previous studies have suggested. All calculations consistently suggest that what is commonly considered to be Ni-redox chemistry predominantly affects the charge state of O (see Figs. 1 a and 4 a). Even though both Ni and O are involved in the redox process, the amount of charge density at the Ni site barely changes on delithiation. If this effect occurred only in a Bader charge analysis, it might be considered an artefact from the choice of Bader cut-offs, not trivial for hybridised states. The DMFT calculations based on a Wannier *d-p* model, however, can separate the *d* and *p* contributions and they confirm the constant charge state of $Ni^{2+}$ and the changing charge state of O from –1.5 in $LiNiO_2$ to –1 in $NiO_2$ (and –2 in NiO). This raises the question of why the charge density changes so much more at the O site than at the Ni site. Examining the Wannier on-site energies allows for an estimation of the energy levels of the localised states. The on-site energies (see Fig. 4 b) show a clear separation between the Ni *d* and O *p* states (Mott-type behaviour) only for the prototype rock salt NiO. In $LiNiO_2$, the difference between the O *p* states and Ni *d* states is < 1eV, promoting hybridisation (as directly seen in the charge density in Fig. 1 c). The strong hybridisation leads to pronounced splitting of the energy levels, pushing O states right up to the Fermi energy (as seen in the DOS in Fig. 1 b), resulting in charge-transfer character. As LiNiO$_2$ is delithiated, the O states are raised in energy (Fig. 4 b). The Ni $e_g\pi$ states and O $p$ states become degenerate in NiO$_2$, maximising the splitting of the hybridised levels. Delithiation thus increasingly pushes the hybridised O states higher in energy, with charge transfer stabilising Ni with a charge of +2 in both LiNiO$_2$ and NiO$_2$. The charge-transfer nature of LNO is also predicted according to the criterion proposed by Zaanen, Sawatzsky, and Allen,[24,25,52] for 3$d$ transition metal compounds, namely that the split between $d$ and $p$ states ($\Delta$ = 2 eV for LNO) is small compared to the $d$–$d$ split (the Coulomb repulsion $U$ = 6 eV). Korotkin et al.[18] and Foyevtsova et al.[19] also speak of ligand holes on oxygen in LNO, discussing the possibility of charge-transfer behaviour of LNO.

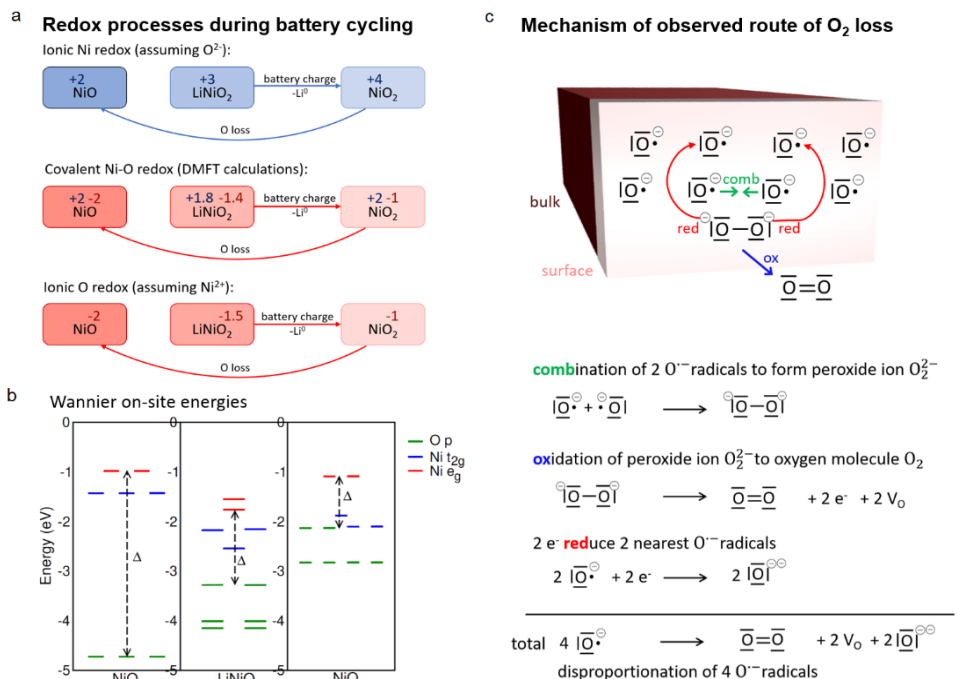

**Fig. 4 Proposed models of the origin and mechanism of O$_2$ loss in LNO. a**, scheme illustrating the differences between the established ionic Ni redox model of LNO and the covalent Ni-O redox observed with DMFT calculations (see Fig. 1) which is very close to an ionic O redox model. The standard understanding considers Ni to undergo the greatest changes in oxidation state when a battery is cycled, from +3 in LiNiO$_2$ to +4 in NiO$_2$ and +2 in the prototype rock salt NiO during degradation. O is commonly considered to be less affected, mostly assuming a charge state of -2. All our DFT and DMFT simulations instead suggest that O undergoes the greatest changes in charge state, from -1.5 in LiNiO$_2$ to -1 in NiO$_2$ and -2 in NiO. Ni is less affected and mostly assumes a charge state of +2. **b**, Wannier on-site energies for NiO, LiNiO$_2$, and NiO$_2$ are a measure of the energies of the localised states. The energies show a small gap Δ between the Ni $d$ and O $p$ states in LNO, promoting Ni–O hybridisation and charge-transfer behaviour. The gap decreases upon delithiation; in NiO$_2$, the Ni $d$ and O $p$ states are degenerate. This causes strong splitting of the levels of the hybridised states, presumably pushing O states close to the Fermi level. This charge-transfer behaviour and O-redox activity directly compromise the material's stability and promote O$_2$ loss. **c,** the following mechanism is derived for the route of spontaneous O$_2$ loss observed in AIMD simulations: (i) 2 oxide radicals O$^{•-}$ combine to form a peroxide ion O$_2^{2-}$; (ii) the peroxide is oxidised to an O$_2$ molecule, desorbs, and leaves behind 2 oxygen vacancies and 2 electrons; (iii) the 2 electrons reduce the closest O$^{•-}$ radicals to O$^{2-}$ ions. In sum, the reaction mechanism thus consists of 4 O$^{•-}$ radicals disproportionating to form molecular O$_2$, 2 O$^{2-}$ ions, and 2 oxygen vacancies. The rate-determining step is the peroxide formation.

Comparing our theoretical XAS Ni $K$ and O $K$-edge spectra to experimental XAS data (see Fig. 2), we see excellent agreement for calculations yielding a constant Ni charge of +2 and varying O charge of −1.5 in LNO and −1 in NiO. Both the observed features and their evolution on delithiation agree very well with experiment, confirming the charge states proposed here. The question then arises why XAS spectra of Ni-rich layered oxides with comparable edge shifts[15,16] have previously been considered evidence of Ni$^{3+}$ changing to Ni$^{4+}$. This calls for a brief methodological reflection on the interpretation of XAS.

XAS is a powerful method for probing local structural environments, both in terms of charge states and bonding geometries. There is, however, to date no way to deconvolute effects from the two, i.e., changes in the XAS edges could arise from changes in the charge state and/or changes in bond lengths, angles etc. Most studies to date interpret the shift of the Ni $K$ edge as evidence of the Ni oxidation state changing.[15] The local coordination environment, however, also changes substantially upon delithiation (see Fig. 2e), mainly in two regards. First, the Jahn-Teller distortions are lifted and, second, leaving aside changes in the symmetry of the bonding environment, the average Ni–O bond lengths also change on delithiation, decreasing from 2.0 Å in LNO to 1.9 Å in NiO$_2$. The shorter bond lengths lead to greater orbital overlap, typically manifesting itself at higher energies in the XAS spectrum. Furthermore, due to the strong Ni-O hybridisation, the Ni $K$ edge is also expected to be altered by changes in the O charge state. The stronger shifts in the O $K$ edge, seen both computationally (Fig. 2b) and experimentally (Fig. 2d) illustrate that the greatest changes in electronic structure occur on O, where in addition to the changing bond lengths the charge state also changes substantially.

Tracing the Ni-edge shift back to changing Ni charge states is therefore by no means a clear-cut deduction; the excellent agreement of the experimental Ni $K$ and O $K$ spectra and our spectra calculated at a constant Ni charge state of +2 on the contrary suggests that it is time to revisit the established interpretation; an assignment of the shift to the changing bond lengths and oxygen charge states is consistent with all our computational and reported experimental data.

It can be concluded that the electrochemical redox activity of LNO stems, to a large extent, from oxidising O, as illustrated in Fig. 4 a. The effect is so pronounced that the redox behaviour of LNO can be described very well with an ionic model after all, but with one phrased in terms of changes of the charge state of O. *N.B.*, we expect O to be strongly involved in many other oxides' redox chemistries, too, but for scenarios where both O and the transition metal undergo significant changes in charge states (*e.g.* in Li-excess materials) it may not be helpful to apply an ionic picture focusing only on one kind of redox-active species.

Identifying the central role of O redox in LNO does not open routes to enhanced capacities as the overall capacity of the hybridised Ni–O system remains constant – and is also limited by the total content of lithium in the material, but it has central ramifications for the material's stability at high states of charge. If bulk Ni–O redox in LNO affects primarily the charge states of oxygen and O is oxidised on battery charging, it is not surprising that the material is unstable towards oxygen loss at high states of charge; and similar behaviour is expected for Ni-rich layered oxide cathodes in general (with the effects increasing with increasing Ni content).

The water-assisted bulk O–O-dimer formation mechanism proposed in Section 2 as a possible origin of the 531 eV peak in the O *K*-edge XAS spectrum at high states of charge can also be rationalised in terms of the pronounced O redox behaviour of LNO. A charge analysis indicates that the dimer itself is oxidised, reducing the nearby lattice O from −1 to *ca.* −1.5, thus counteracting the oxidation of bulk O. An in-depth study will follow shortly. The dimers cause XAS signal intensity around 531 eV, where features are conventionally assigned to O-redox processes[53] (and they exhibit an O–O distance of *ca.* 1.2 Å, consistent with the vibronic features seen with Resonant Inelastic X-ray Scattering, RIXS[17,53]). As the dimer formation occurs because of O redox, the 531 eV XAS feature can technically be considered an O-redox signature. Given that O redox in our eyes governs the evolution of the whole spectrum on delithiation, there is, however, little benefit in highlighting specifically the 531 eV signal as an O-redox feature, as the changes in edge energies and intensities are all a result of the oxidation of the lattice O on delithiation. We leave for a future study to clarify further why the 531 eV peak emerges at high voltages: specifically whether it requires either a critical extent of O oxidation, of electrolyte decomposition generating water, or of free space for the water molecule to enter the Li layer – or most likely a combination of these factors. Our findings further open up a bigger discussion regarding the O-redox feature reported for various materials at high states of charge, ranging from the NMCs to Li-excess materials.[53] Given that these materials differ substantially in many regards but are all oxides, the question arises if the XAS peak often assigned to O redox in some of these materials can be traced back to water reacting with oxidised lattice O to form O–O dimers. We believe this is a likely scenario which warrants a broad investigation, including a discussion of conceivable alternative origins of the high-voltage peak such as electronic quasi-particles (see SI).

Methodologically, we find PBE+$U$ calculations with $U_{eff}$ = 6 eV surprisingly well suited for describing the 0 K electronic structure – particularly with respect to the charge states of LNO, $NiO_2$, and NiO (see Fig. 1 a). DMFT calculations based on a Wannier $d$–$p$ model not only validate the trends seen with a PBE+$U$ Bader charge analysis, and suggest they remain valid at finite temperatures, but also show excellent quantitative agreement with PBE Wannier charges. The wannierisation route for determining charges in hybridised states[28,29] in our case obtained DMFT-quality oxidation states from PBE when the basis of the Wannier model was chosen carefully ($d$–$p$ for $LiNiO_2$ and $NiO_2$; $d$ for NiO). It promises to be a powerful tool for quantifying charge states in battery materials more generally.

Based on our AIMD simulations of spontaneous $O_2$ loss from the (012) facet in $NiO_2$ (see Fig. 3), we propose the following mechanism for the observed reaction route: (i) 2 oxide radicals $O^{•-}$ combine to form a peroxide ion $O_2^{2-}$ (see Fig. 4 c); (ii) upon peroxide formation and desorption, the 2 $O^{•-}$ are oxidised to molecular $O_2$, leaving behind 2 oxygen vacancies and 2 electrons; (iii) the 2 electrons reduce the 2 closest $O^{•-}$ radicals to $O^{2-}$ ions. (ii) and (iii) occur simultaneously. In sum, the reaction mechanism thus consists of 4 $O^{•-}$ radicals disproportionating to form molecular $O_2$, 2 $O^{2-}$ ions, and 2 oxygen vacancies. Whereas in metal-air batteries, the superoxide radical $O_2^{•-}$ disproportionates,[30] the reaction route we observe in LNO involves the disproportionation of the oxide radical $O^{•-}$, *i.e.*, while there are parallels between the reaction mechanisms, the disproportionating species differ according to the different battery redox chemistries. The oxide $O^{•-}$ disproportionation is likely to be representative of Li-ion batteries with Ni-rich layered oxide cathodes.

After oxygen is lost, Ni migration into the Li layer is expected to result in the formation of Ni-densified surface phases (rock-salt phases).[1] We leave a more detailed investigation of the densification process for future study but point out that both Ni migration and O loss in Ni-rich oxide cathodes occur due to the thermodynamic instability of the delithiated phase. We here identify the origin of the instability of $NiO_2$ (and the related phenomena) to lie in the significant covalency of $LiNiO_2$ resulting in an apparent high oxidation state of O (–1) in $NiO_2$. A further point is the question of whether densified surface phases prohibit the further release of oxygen. While the rock-salt phases most likely passivate the surfaces, either prohibiting or slowing down further oxygen loss, polycrystalline Ni-rich oxides are known to suffer from particle cracking[54] which continues to expose fresh surfaces: until particle cracking can be successfully prevented, the materials are expected to continue releasing oxygen via the reported and related reaction pathways.[55-57]

The reason for the formation of singlet oxygen, we believe, lies in the ground-state singlet state of the peroxide ion. As the peroxide is oxidised to molecular oxygen, spin conservation rules favour the release of $^1O_2$. Given the strongly exergonic nature of the reaction (see Fig. 3 b) liberating 3 eV, $^1O_2$ is easily released in its excited state. Singlet oxygen formation in layered oxide cathodes thus follows in the footsteps of the wealth of reports of singlet oxygen formation from (mostly organic) peroxide reactions.

The final question concerns the fate of $^1O_2$ after its evolution. Does it remain in the singlet state (accounting for $^1O_2$ found experimentally[7]), react with electrolyte components (explaining the vast amounts of electrolyte oxidation products found), or deactivate into its triplet ground state? Both during the reaction and shortly after, the peroxide/$O_2$ molecule strongly interacts with the spins at the surface. Surface interactions, for example, enable the combination of the two $O_2^{\bullet-}$ radicals, each with positive spin density, to the singlet peroxide in the first place – a process that would be spin-forbidden for isolated radicals. Interactions of the newly formed singlet $^1O_2$ molecule with the surface will also provide routes of the otherwise spin-forbidden relaxation into triplet $^3O_2$. We leave these processes for future study, ideally to be addressed with dedicated multi-reference tools that can fully account for the open-shell character of the isolated $O_2$ molecule and excited states beyond the DFT ground state. Our work, however, shows that a low activation energy route towards singlet oxygen release occurs via the ground state singlet state of the peroxide.

# 6 Conclusion

Through a rigorous charge analysis based on DFT and DMFT calculations, we have shown that the delithiation of $LiNiO_2$ to $NiO_2$ predominantly occurs via the oxidation of O and only causes minor changes in the charge of Ni. The Ni charge remains around +2, whether in $LiNiO_2$, in $NiO_2$, or the prototype rock salt NiO. Instead, the oxygen charge ranges from −2 in NiO to *ca.* −1.5 in $LiNiO_2$ and −1 in $NiO_2$, and we propose it is the charge-transfer character of LNO and the resulting dominant O redox that cause oxygen loss in Ni-rich layered cathode materials. Calculated XAS Ni *K* and O *K*-edge spectra are in excellent agreement with experimentally obtained spectra, confirming our charge analysis. A high-voltage feature commonly assigned to O redox could be accounted for by considering reactions of intercalated water molecules, which can combine with lattice O to form O–O dimers at high states of charge. Our AIMD simulations of the (012) facet of $NiO_2$ show the spontaneous evolution of $O_2$ molecules and the occurrence of singlet oxygen in the process. We proposed the following mechanism for the observed route of $O_2$ loss: (i) 2 $O^{\bullet-}$ radicals combine to form a peroxide ion $O_2^{2-}$; (ii) the peroxide ion is oxidised to molecular $O_2$ which then leaves the surface, leaving behind 2 oxygen vacancies and 2 electrons; (iii) the remaining 2 electrons reduce the 2 nearest lattice $O^{\bullet-}$ radicals to $O^{2-}$, steps (ii) and (iii) occurring simultaneously. Overall, this means 4 $O^{\bullet-}$ radicals disproportionate to form $O_2$ and 2 $O^{2-}$ ions. The reaction liberates *ca.* 3 eV. Singlet oxygen formation is caused by the singlet ground state of the peroxide ion. Spin conservation favours the release of $^1O_2$ in the excited state, which is rendered feasible through the strongly exergonic nature of the reaction. The findings presented here not only propose a revised model of redox processes in stoichiometric LNO but also offer a comprehensive atomistic understanding of the cause of oxygen loss and a reaction route forming singlet oxygen. Mitigating $O_2$ loss in Ni-rich cathodes and designing more reversible high-voltage cathode materials thus requires a shift in the understanding of transition metal-oxygen redox. Only if the pivotal role of O redox is acknowledged, can routes be sought to suppress oxidation – especially of the surface O, *e.g.* through adding dopants that promote more metal-centered redox rather than O-centered redox; and routes preventing $O_2$ release in well-known O-redox materials[58] can be explored for mitigating (singlet) oxygen release from Ni-rich layered oxides.

# 7 Methods

**AIMD simulations and DFT calculations**

AIMD simulations and static DFT calculations were performed according to the Generalized Gradient Approximation (GGA) proposed by Perdew, Burke, and Ernzerhof[59] and the projector augmented wave method(PAW),[60] as implemented in the Vienna Ab Initio Simulation Package (VASP).[61,62] The electronic wave functions were expanded with a basis set of plane waves with kinetic energies of up to 500 eV. For the charge calculations, supercells with 32-64 ions and a 7 x 7 x 7 Monkhorst-Pack *k*-point mesh[63] were used. The convergence criteria for the electronic and ionic relaxations were set to $10^{-8}$ eV and $1 \cdot 10^{-3}$ eV/Å, respectively. For the MD simulations, a supercell with 48 ions was used. The *k*-point sampling was based on a 1 x 2 x 1 Monkhorst-Pack mesh[63]. The convergence criteria for the electronic and ionic relaxations were set to $10^{-6}$ eV and $5 \cdot 10^{-3}$ eV/Å, respectively.

For Ni, the $4s^2 3d^8$ electrons were treated as valence electrons. To account for the strongly correlated *d* electrons, a rotationally invariant Hubbard *U* parameter was used.[64] The electronic density of states was calculated at varying $U_{eff}$ in the range from 0 eV – 10 eV and compared with the density of states obtained with the screened hybrid functional proposed by Heyd, Scuseria, and Ernzerhof (HSE06)[65] with 25% Fock exchange. The best agreement was achieved at $U_{eff}$ = 6 eV, in agreement with findings by Das *et al.*,[5] which was used for all AIMD and DFT calculations. For oxygen, the $2s^2 2p^4$ electrons were considered in the valence.

AIMD simulations were performed for the canonical ensemble (*NVT*, constant volume, particle number, and temperature). A Nosé-Hoover thermostat was used with a Nosé mass corresponding to a period of 40 fs.

**Dynamical Mean-Field Theory**

Our *ab-initio* dynamical mean-field theory (DMFT) calculations are based on the full-potential augmented plane-wave basis as implemented in WIEN2K.[66] For these calculations, we used the largest possible muffin-tin radii, and the basis set plane-wave cut-off as defined by $R_{min} \cdot K_{max} = 8$, where $R_{min}$ is the muffin-tin radius of the oxygen atoms. The consistency between VASP and WIEN2K results has been cross-checked.

We perform the DMFT calculations in a basis set of Maximally Localised Wannier Functions (MLWF) using Wannier90[67] and the wien2wannier[68] interface. DMFT calculations were performed using the TRIQS/DFTTools package[69-71] based on the TRIQS libraries.[72] Projective Wannier functions using the DMFTproj module of TRIQS/DFTTools were used to calculate the initial occupancies of the correlated orbitals. DMFT calculations in both MLWF and projective Wannier function (DMFTproj) basis was found to yield consistent results. In both cases a projection window of -8 eV to +2 eV was chosen. For both paramagnetic and magnetic calculations, Ni *d* and O *p* orbitals have been considered for the DMFT calculation, since the O *p* orbitals are higher in energy and closer to Ni *d* orbitals and are, thus particularly important in case of LiNiO$_2$ and NiO$_2$ where a significant charge transfer is expected. Calculation of *d* and *p* occupancies using DMFTproj in this energy range yields significantly different occupancies from the expected formal oxidation states for each element which is adjusted only slightly by DMFT calculations on the *d* manifold with projections for O *p* calculated and made part of the bath.

It is to be noted that Ni *d* states strongly hybridise with O *p*. This hybridisation may also in general be taken into account by considering a downfolded *d only* orbital basis, in which O *p* degrees of freedom are not thrown away but included in the tail of the *d* Wannier functions (as shown in Fig. 1). However, this excludes any charge transfer possibility between Ni *d* and O *p* orbitals, which is imminent in case of LiNiO$_2$ and NiO$_2$ since the *d*–*p* split $\Delta$ is smaller than or at best equal to U and should facilitate charge transfer. It is also to be noted that due to the large *d*–*p* split $\Delta$ a simple downfolded *d only* orbital basis is sufficient to describe NiO. In both NiO$_2$ and LNO due to the presence of trigonal distortion in the NiO$_6$ octahedra, the t$_{2g}$ orbitals are not degenerate, rather they are split into singly degenerate a$_{1g}$ and doubly degenerate e$_g\pi$. The energy difference between a$_{1g}$ and e$_g\pi$ is small, hence may be treated as almost degenerate. Wannier90 projects to a global coordinate system, and hence any projections with spherical harmonics correctly identify the orbitals in their projections.

The Anderson impurity problems were solved using the continuous-time quantum Monte Carlo algorithm in the hybridisation expansion (CT-HYB)[73] as implemented in the TRIQS/CTHYB package.[74] We performed both one-shot and fully charge self-consistent calculations, with the double-counting correction treated in the fully-localised limit.[38] A good agreement of both methods is seen. We used the full rotationally-invariant version of the Kanamori interaction.[75] For our calculations we used $U_{dd}$ values ranging from 4.5 – 8 eV and $J$ varying between 0.5 - 0.75 eV to investigate the whole spectrum of metal-to-insulator transition. We set the intra-orbital interaction to $U' = U - 2J$.

Real-frequency results have been obtained using the maximum-entropy method of analytic continuation of the Green's functions as implemented in the TRIQS/MAXENT module.[76]

**Prediction of X-ray absorption spectra**
X-ray absorption spectra of the O $K$ edge were calculated with the plane-wave code CASTEP using norm-conserving pseudopotentials with an energy cutoff of 1500 eV. A core hole was introduced into the 1$s$ state.[77] The spectra were evaluated with the OptaDOS software package with the zero of each spectrum set to the pre-peak onset.

The X-ray absorption spectra of transition metals are commonly dominated by atomic multiplet interactions arising from electron correlation.[78] The Ni $K$-edge spectra were therefore calculated with the FEFF 10 code which explicitly considers multiplet interactions.[79,80] FEFF employs Green's formulation of the multiple scattering theory to compute the spectra. The X-ray absorption $\mu$ is calculated in a manner similar to Fermi's golden rule when written in terms of the projected photoelectron density of final states or the imaginary part of the one-particle Green's function, G($r,r'$;E). In terms of the Green's function, G($r,r'$;E), the absorption coefficient, $\mu$, from a given core level **c** is given by[81]

$$\mu = -\frac{1}{\pi} \text{Im} <c| \; \varepsilon \cdot r \; G(r,r';E) \; \varepsilon \cdot r' \; |c>$$

with the Green's function, G($r,r'$;E) given by

$$G(r,r';E) = \sum_f \frac{\Psi_f(r)\Psi_f^*(r\prime)}{E - E_f + i\Gamma}$$

where $\Psi_f$ are the final states, with associated energies $E_f$ and net lifetime $\Gamma$, of a one-particle Hamiltonian that includes an optical potential with appropriate core-hole screening. The FEFF code computes the full propagator G incrementally using matrix factorisation and uses the spherical muffin-tin approximation for the scattering potential.[81] For self-consistent potential calculations required in the XANES calculation for the Fermi level $E_0$ estimation a large value of rfms1 of 9 Å was chosen to have a large number of atoms included in the self-consistent potential calculations. *Full multiple scattering (FMS)* is required in the XANES calculation as the multiple scattering (MS) expansion's convergence might not be stable in the XANES calculation. A large rfms value of 11 Å was considered for proper convergence. The Hedin-Lundqvist self-energy was chosen for the exchange correlation potential model used for XANES calculations. The random phase approximation (RPA) is used to approximate the core-hole interactions in our Ni *K*-edge *XANES* calculations.

## Acknowledgements

This work was supported by the Faraday Institution degradation project (FIRG011, FIRG020). A. R. Genreith-Schriever gratefully acknowledges funding from the German National Academy of Sciences Leopoldina. We thank Angela Harper and Rebecca Nicholls for fruitful discussions. Generous computing resources were provided by the Sulis HPC service (EP/T022108/1), the University of Birmingham's BlueBEAR HPC service, and networking support by CCP-NC (EP/T026642/1), CCP9 (EP/T026375/1), and UKCP (EP/P022561/1).